\documentclass[pra,twocolumn,showpacs]{revtex4}
\usepackage{amsfonts}
\usepackage{amsmath}
\usepackage{amssymb}
\usepackage{graphicx}
\usepackage{epstopdf}

\begin{document}

\title{Stability of Z$_2$ topological order in the presence of vacancy-induced impurity band}
\author{Shi-Ting Lee$^1$, Shin-Ming Huang$^{1}$, and Chung-Yu Mou$^{1,2,3}$ }
\affiliation{$^{1}$Department of Physics and Frontier Research Center on Fundamental and
Applied Sciences of Matters, National Tsing Hua University, Hsinchu 30043,
Taiwan}
\affiliation{$^{2}$Institute of Physics, Academia Sinica, Nankang, Taiwan}
\affiliation{$^{3}$Physics Division, National Center for Theoretical Sciences, P.O.Box
2-131, Hsinchu, Taiwan}

\begin{abstract}
Although topological insulators (TIs) are known to be robust against
non-magnetic perturbations and exhibit edge or surface states as their
distinct feature, experimentally it is known that vacancies often occur in
these materials and impose strong perturbations. Here we investigate effects
of vacancies on the stability of Z$_{2}$ topological order using the
Kane-Mele (KM) model as a prototype of topological insulator. It is shown
that even though a vacancy is not classified as a topological defect in KM
model, it generally induces a pair of degenerate midgap states only in the
TI phase. We show that these midgap states results from edge states that fit
into vacancies and are characterized by the same Z$_{2}$ topological order.
Furthermore, in the presence of many vacancies, an impurity band that is
degenerate with edge states in energy is induced and mixes directly with
edge states. However, the Z$_{2}$ topological order persists and edge states
exist between the impurity band and perturbed bulk bands until a phase
transition occurs when Dirac cones near Dirac points are depleted. Our
analyses indicate that the same scenario holds for point vacancies or line
of vacancies in 3D TIs as well.
\end{abstract}

\maketitle

\section{Introduction}

Building low dimensional electronic systems has been a focus of intense
interest since the discovery of low dimensional materials such as carbon
nanotubes and graphene. Recent theoretical and experimental works on
topological order of materials \cite{Kane,Hasan,TI} have led to the
discovery of a different route to construct low dimensional electronic
systems through a new class of materials, called topological insulators
(TIs). In these materials, surface states or edge states arise in the
boundaries of bulk insulating materials and can host electrons in low
dimensions. Depending on the dimension of the materials, a two-dimensional
TI hosts one-dimensional gapless edge states while a three-dimensional TI
hosts two-dimensional gapless Dirac Fermions. These low dimensional surface
states or edges arise from Z$_{2}$ topological order of bulk states \cite%
{Kane} and are protected by symmetries of bulk states.

Theoretically, the fundamental reason for the emergence of surface states or
edge states in TI is due to underlying Z$_{2}$ topological structure in the
bulk state. Since topological order is protected by associated symmetries
\cite{symmetry}, surface states and edge states in TIs are considered to be
robust against bulk disorders that do not break the time-reversal symmetry
that is associated with Z$_{2}$ topological order. The robustness against
disorders is the key to potential technological applications and hence a
great effort has been dedicated to understanding the behavior of the
topological materials in the presence of disorders \cite{Sheng1, disorder1,
disorder2, disorder3, 3Dstrong, disorderShen, Beenakker}. Indeed, surface
states or edge states are shown to persist in the presence of weak disorders
\cite{Sheng1}. However, it is also found that in the presence of disorders,
anomalous transmutation between TI and insulators with trivial topology may
happen \cite{3Dstrong,disorderShen,Beenakker}. In particular, in computer
simulations of a HgTe quantum well, it is discovered that an ordinary
insulating state can be transformed into a topological insulator, called
topological Anderson insulator \cite{disorderShen, Prodan}. The mechanism
behind the transformation is shown to be related to the renormalization of Z$%
_{2}$ topological index by disorders near the Dirac points \cite{Beenakker}
and the transformation may also depend on type of disorders \cite{Xie}.
Hence while TI is robust against perturbations, the exact phase boundary for
TI may subject to change in the presence of disorders.

Experimentally, it is found that strong perturbations such as lattice
defects often occur in topological insulators \cite{exptvacancy}. From the
experience in other similar materials characterized by Dirac Hamiltonian,
the presence of lattice defects often induces peculiar properties. In the
case of graphene, it was found that vacancies can give rise to magnetic
moments and turn graphene into a ferromagnet \cite{mou}. For topological
insulators, it is shown that lattice defects with non-trivial topology also
induce bound states \cite{lattice}. However, in the tetradymite
semiconductor Bi$_{2}$Se$_{3}$ that has been most extensively investigated,
the most common observed lattice defects are selenium vacancies, which are
not topological defects. These vacancies are believed to give rise
to electron doping and increase the conductivity of bulk states dramatically
\cite{exptvacancy}. Furthermore, there are also evidences that vacancies may
induce an impurity band that affects current transportation \cite{Ando}.
Since vacancies do not break time-reversal symmetry, their presence is
consistent with symmetries that are associated with Z$_{2}$ topological
order. However, instead of being weak, vacancies are considered as strong
perturbations and may change the topological structure of TI. It is
therefore interesting and crucial to examine robustness of TI in the
presence of vacancies.

In this paper, we will examine the stability of Z$_{2}$ topological order in
the presence of vacancies using the Kane-Mele model. It is shown that even
though vacancies are not classified as topological defects in the Kane-Mele
model \cite{defect}, the original Z$_{2}$ topological order insures that
midgap bound states are induced only in topologically nontrivial phase.
Based on Green's function analysis of edge states \cite{mou1}, we show that
these midgap bound states are remanent edge states and form an impurity band
in the presence of many vacancies. We shall show that the impurity band
coexists with edge states when the Z$_{2}$ topological order persists until
the spectral weights of Dirac cones near Dirac points are depleted. Finally,
we briefly extend our analyses to three dimensions and show that there must
be midgap bound states associated with point vacancies or line of vacancies
in 3D TIs as well.

\section{ Theoretical formulation and midgap bound states near a vacancy}

We start with the Kane-Mele model on a honeycomb lattice. The Hamiltonian is
given by \cite{Kane}
\begin{eqnarray}
H &=&-t\sum_{<i,j>}c_{i}^{\dag }c_{j}+i\frac{\lambda _{SO}}{3\sqrt{3}}%
\sum_{<<i,j>>}\nu _{ij}c_{i}^{\dag }\sigma ^{z}c_{j}  \nonumber \\
&+&i\frac{2\lambda _{R}}{3}\sum_{<i,j>}c_{i}^{\dagger }(\mbox{{\boldmath$%
\sigma$}}\times {\hat{\mathbf{d}}}_{ij})_{z}c_{j}  \nonumber \\
&&+\lambda _{v}\sum_{i}\mu _{i}c_{i}^{\dag }c_{i}+\sum_{i}V_{i}c_{i}^{\dag
}c_{i},
\end{eqnarray}%
where $c_{i}^{\dag }=(c_{i\uparrow }^{\dag },c_{i\downarrow }^{\dag })$
creates an electron at lattice site $i$. The first term accounts for
nearest-neighbor hoppings. The second term is a spin-orbit interaction
between next nearest neighbors, in which $\nu _{ij}=(2/\sqrt{3})(\hat{d_{1}}%
\times \hat{d_{2}})\cdot \hat{z}$ with $\hat{d_{1}}$ and $\hat{d_{2}}$ being
two nearest-neighbor bonds that connect site $j$ to site $i$, and $%
\mbox{\boldmath{$\sigma$}}$ are the Pauli matrices. The third term is the
Rashba coupling. The fourth term characterizes the sublattice site energies
with $\mu _{i}=+1$ $(-1)$ for $i\in A$ ($B$) sites. This term will be
considered only in this section. The last term is the random disorder
potential that models the impurities, where $V_{i}$ takes a nonvanishing
value of $V$ only at impurity sites. In the usual treatment of disorder
potential, it often assumes that $V_{i}$ is smooth and is characterized by
its correlation function $\langle V_{i}V_{j}\rangle $. Here, however, by
taking $V_{i}\rightarrow \infty $, $i$ site is excluded for electrons to
visit and hence the limit $V_{i}\rightarrow \infty $ simulates a vacancy at $%
i$ site. As we shall see, the new ingredient of the lattice vacancy is the
possibility of inducing bound states near a vacancy site, which can not be
obtained perturbatively by using $\langle V_{i}V_{j}\rangle $. Note that
vacancies can be also modeled by cutting all relevant couplings to sites of
vacancies and it can be shown that this approach yields the same results in
low energy sectors where energy bands occupy. In the following, all energies and lengths will be in units of $t$ and $a$
(lattice constant) respectively.

\begin{figure}[tbp]
\begin{center}
\includegraphics[height=2.5in,width=4.0in] {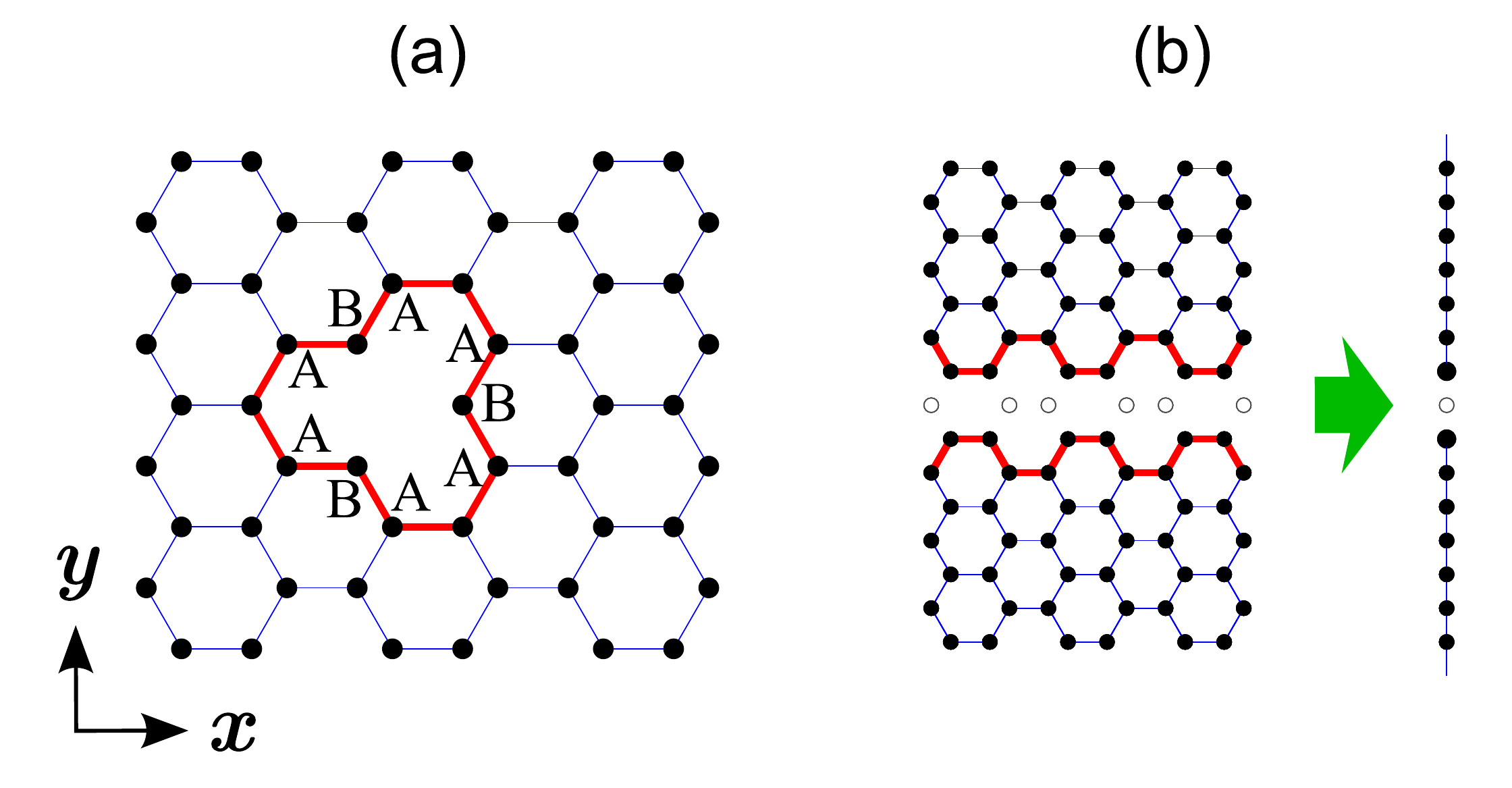}
\end{center}
\caption{(a) A single vacancy locates at the origin $\vec{r}=(0,0)\equiv
\mathbf{0}$, which belongs to $A$ sublattice. (b) A line of point vacancies
that are marked by open circles cut an infinite honeycomb lattice into two
semi-infinite honeycomb lattices with two edges marked by red lines. After
partial Fourier transformation, for each Fourier mode, the system is
dimensionally reduced to an effective vacancy in one dimension. Note that
curving one of the edges can form a vacancy shown in (a), indicating that
mid-gap bound states near a vacancy can be formed by suitable superposition
of edge states.}
\label{fig1}
\end{figure}
We shall first consider the case of a single vacancy located at the origin $%
\vec{r}=(0,0)\equiv \mathbf{0}$ that is located at $A$ site. The
configuration is shown in Fig.~\ref{fig1}(a). According to the
classification made Ref. \cite{defect}, far from a vacancy in 2D, there is
no non-trivial winding number associated with the hole introduced by a
vacancy and hence a vacancy is not a topological defect. However, this does
not imply that there is no midgap bound state associated with a vacancy. In
fact, in the continuum limit for HgTe where the Dirac point is at $\Gamma $
point, by approximating a vacancy as a circular hole with radius $R$, it is
found that midgap bound states survive with energies being half of the gap
magnitude in the limit $R\rightarrow 0$ \cite{vacancyShen}. In the continuum
limit, electrons that occupy the midgap bound states can be viewed as Dirac
fermions in a particular curved space \cite{Lee}. Hence a midgap bound state
in a single vacancy results from a superposition of edge states that are
curved into the circular surrounding of the vacancy. However, for a vacancy
in a honeycomb lattice, the situation is quite different. As shown in Fig.~%
\ref{fig1}(a), a vacancy has a finite size and its surrounding is not in a
circular shape. In addition, Dirac points are located at finite wave vectors
for honeycomb lattices. It is necessary to examine whether there are midgap
bound states associated with vacancies.

To clarify the issue of whether midgap bound states will be induced by a
single vacancy, we resort to method of the Green's function, which has the
advantage of not being restricted by finite size effects. In the presence of
impurities or vacancies, the Green's function $G_{\alpha \beta }({\vec{r},%
\vec{r}^{\prime },E})$ that describes the amplitude for the electron of
energy $E$ to propagate from $\vec{r}^{\prime }$ with spin component $\beta $
to the position $\vec{r}$ with spin component $\alpha $ satisfies
\begin{equation}
\left( E-H\right) G({\vec{r},\vec{r}^{\prime },E})=\delta _{\vec{r},\vec{r}%
^{\prime }},  \label{Geq}
\end{equation}%
where we have collectively represented $G_{\alpha \beta }$ by the $2\times 2$
matrix $G$. If we denote the Green's function in the absence of vacancies by
$g({\vec{r},\vec{r}^{\prime },E})$, we find that for a single impurity at $%
\vec{r}=\mathbf{0}$, $G$ satisfies
\begin{eqnarray}
G({\vec{r},\vec{r}^{\prime },E}) &=&g({\vec{r},\vec{r}^{\prime },E})+g({\vec{%
r},\mathbf{0},E})VG({\mathbf{0},\vec{r}^{\prime },E})  \nonumber \\
&=&g({\vec{r},\vec{r}^{\prime },E})+g({\vec{r},\mathbf{0},E})Tg({\mathbf{0},%
\vec{r}^{\prime },E}),  \label{G}
\end{eqnarray}%
where $T$ is the T-matrix for a single vacancy at $\vec{r}=\mathbf{0}$ and
can be written as
\begin{equation}
T=V+VgV+VgVgV+\cdots =V\left( 1-gV\right) ^{-1}.  \label{Tmatrix}
\end{equation}%
If $H$ supports midgap bound states, $G$ must contain energies of the midgap
bound states as poles in $E$. Therefore, the energies of midgap bound
states, $E_{0}$, are determined by \cite{mou}
\begin{equation}
\left. \det \left( 1-gV\right) \right\vert _{E=E_{0}}=0.  \label{det}
\end{equation}%
In the limit of $V\rightarrow \infty $, the impurity becomes a vacancy. Eq. (%
\ref{det}) reduces to equations that determine $E_{0}$ for midgap bound
states of vacancies
\begin{eqnarray}
\det g(\mathbf{0},\mathbf{0},E_{0}) &=&g_{11}\left( \mathbf{0},\mathbf{0}%
,E_{0}\right) g_{22}\left( \mathbf{0},\mathbf{0},E_{0}\right)   \nonumber \\
&-&g_{12}\left( \mathbf{0},\mathbf{0},E_{0}\right) g_{21}\left( \mathbf{0},%
\mathbf{0},E_{0}\right) =0,  \label{Ebound}
\end{eqnarray}%
where $1$ and $2$ represent spin up and down respectively. We shall show in
below that the existence of a midgap bound state for a single vacancy
results from the existence of edge states.

To connect edge states with midgap bound states of a single vacancy, we
shall start from a system that is a $Z_{2}$ topological insulator. Hence
there are helical edges states for any edges. Furthermore, the energy
spectrum of edge states is a Dirac-like spectrum \cite{Kane}. For a given
edge along $\zeta $-direction, since it is translationally invariant along $%
\zeta $-direction, the system can be dimensionally reduced to one dimension
by a partial Fourier transformation along $\zeta $-direction. Let the wave
vector along $\zeta $ be $k_{\zeta }$. The existence of edge states thus
implies that there are states with energies $E$ being inside the bulk energy
gap and being in the form \cite{Kane}
\begin{equation}
E=\mu _{\zeta }\pm v_{\zeta }k_{\zeta },  \label{helical}
\end{equation}%
where $v_{\zeta }$ is the speed of helical edge states and $\mu _{\zeta }$
is the intersecting energy of two helical modes (energy of the Dirac point).
Note that the effective Hamiltonian that gives rise to the helical spectrum
of Eq. (\ref{helical}) is a Dirac Hamiltonian and can be written as \cite%
{Kane}
\begin{equation}
H_{eff}=\left(
\begin{array}{cc}
\mu _{\zeta } & v_{\zeta }k_{\zeta } \\
v_{\zeta }k_{\zeta } & \mu _{\zeta }%
\end{array}%
\right) ,  \label{heff}
\end{equation}%
where $H_{eff}$ is a $2\times 2$ matrix in the spin space. In the following,
we shall show that a midgap bound state for a single vacancy is essentially
a weighted superposition of edge states of all possible orientations
(denoted by $\zeta $) characterized by Eq. (\ref{helical}).

We now consider creating edges by introducing a line of point vacancies in
an infinite honeycomb lattice as illustrated in Fig.~\ref{fig1}(b). The
vacancy line cuts the infinite honeycomb lattice into two semi-infinite
honeycomb lattices with two edges. As we shall show in below, after partial
Fourier transformation on coordinates along the vacancy line, for each
Fourier mode $k_{\zeta }$, the vacancy line becomes a point. That is,
through dimensional reduction, the vacancy line and edge states that are
associated with two edges reduce to an effective vacancy with midgap bound
states in one dimension. The connection of midgap bound states for a single
vacancy to edge states is thus established. From this point of view, we may
assume that the vacancy line passes through $\vec{r}=\mathbf{0}$ and its
direction is denoted as $\zeta $-direction. Hence the resulting edge for
each semi-infinite honeycomb lattice is along $\zeta $-direction. For later
usage, we denote the direction perpendicular to $\zeta $ by $\eta $ with the
corresponding coordinate being denoted by $x_{\eta }$. In the example shown
in Fig.~\ref{fig1}(b), one has $\zeta =x$, $\eta =y$, and $k_{\zeta }=k_{x}$%
. After the partial Fourier transformation along $\zeta $-direction, the
Hamiltonian of the whole system (two semi-infinite honeycomb lattices plus a
vacancy line) is reduced to one-dimensional subsystems of different $%
k_{\zeta }$, hence the total Hamiltonian $H$ can be written as
\begin{equation}
H=\sum_{k_{\zeta }}h(k_{\zeta }).
\end{equation}%
Note that after the partial Fourier transformation, the vacancy line reduces
to an effective vacancy for each $k_{\zeta }$, and each $h(k_{\zeta })$
contains a vacancy at the origin of \textit{A} site, $x_{\eta }=0$. The
associated Green's function is a function of $k_{\zeta }$ and $x_{\eta }$
and can be written as $G_{\alpha \beta }({x_{\eta },x_{\eta }^{\prime
},k_{\zeta },E})$. Clearly, following the derivation that leads to Eqs. (\ref%
{G}) and (\ref{Tmatrix}), since there is a vacancy at $x_{\eta }=0$, we find
that for a given $k_{\zeta }$, the T-matrix is given by
\begin{equation}
T(k_{\zeta },E)=V\left( 1-g(x_{\eta }=0,x_{\eta }^{\prime }=0,k_{\zeta
},E)V\right) ^{-1}.  \label{Tmatrix1}
\end{equation}%
In the limit of $V\rightarrow \infty $, one gets
\begin{equation}
T=-\left[ g(x_{\eta }=0,x_{\eta }^{\prime }=0,k_{\zeta },E)\right] ^{-1}.
\label{Tmatrix2}
\end{equation}

Since edge states are created by the vacancy line, according to Eq. (\ref%
{helical}), $\mu _{\zeta }\pm v_{\zeta }k_{\zeta }$ must be eigen-energies
of $h(k_{\zeta})$ and hence these energies must appear as poles of the
T-matrix. In addition, using the fact that edge states are described by the
effective Hamiltonian given in Eq. (\ref{heff}) and their energies must be
poles of $T$, it implies that
\begin{equation}
T \sim (E-H_{eff})^{-1}.  \label{Tmatrix3}
\end{equation}
As a result, by combining Eqs. (\ref{Tmatrix2}) and (\ref{Tmatrix3}), we
conclude that $g$ will take the following form
\begin{equation}
g(x_{\eta }=0,x^{\prime }_{\eta }=0,k_{\zeta },E)=w(k_{\zeta },E)\left(
\begin{array}{cc}
E-\mu _{\zeta } & v_{\zeta }k_{\zeta } \\
v_{\zeta }k_{\zeta } & E-\mu _{\zeta }%
\end{array}%
\right) ,  \label{1Dg}
\end{equation}
where the factor $w(k_{\zeta },E)$ is a proportional constant and accounts
for the weight associated with $k_{\zeta}$ mode. Note that $w(k_{\zeta },E)$
is smooth, even in $k_{\zeta}$ and it has no zeros in $E$.
\begin{figure}[tbp]
\begin{center}
\includegraphics[height=3.1in,width=3.75in] {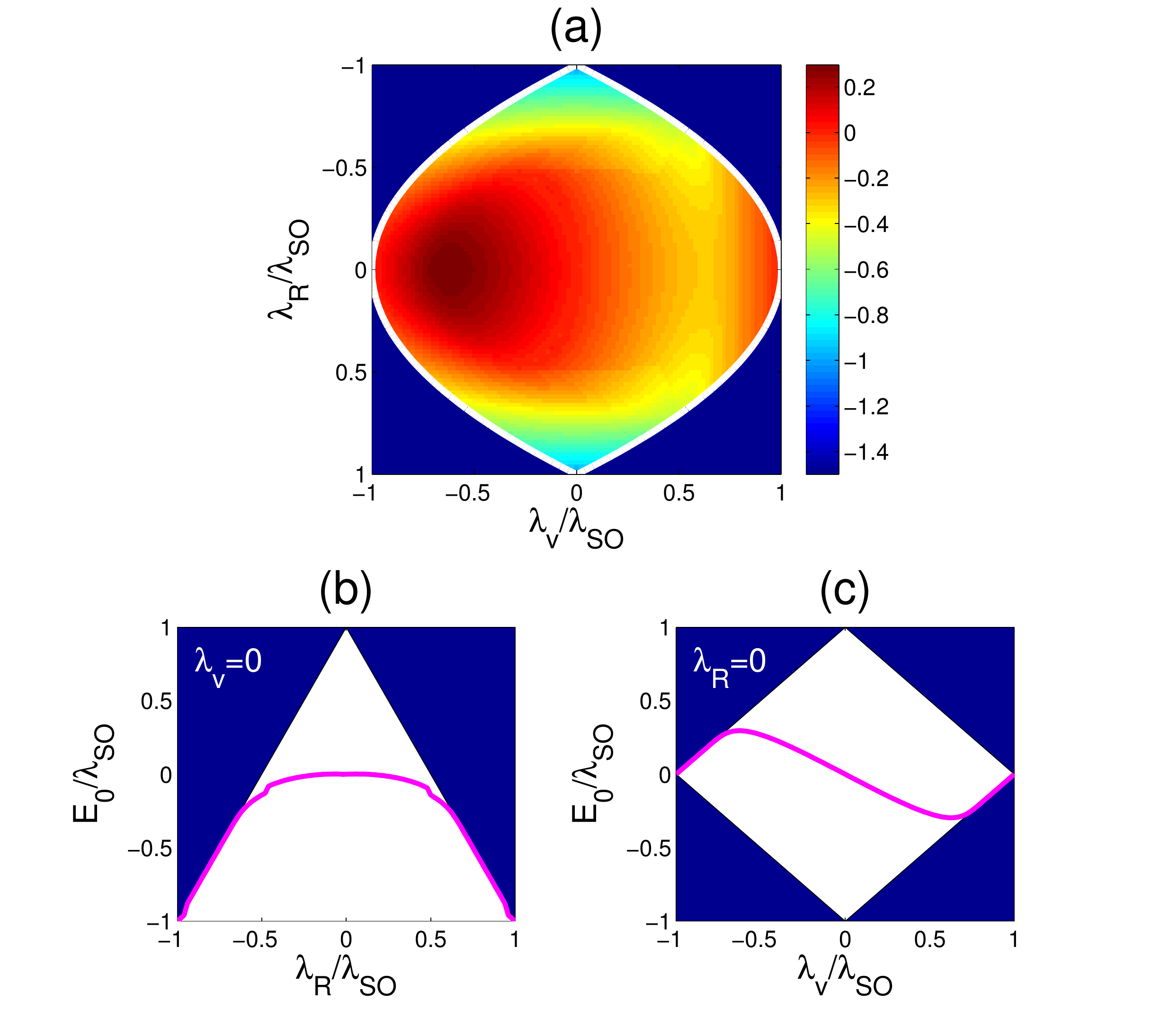}
\end{center}
\caption{(Color online) Energies ($E_0$) of mid-gap bound states induced by
a single vacancy. (a) Color map of $E_0/\protect\lambda _{SO}$. Mid-gap
bound states only appear inside the TI phase. The region outside the TI
phase (enclosed by the white line) is marked as blue to indicate that there
is no mid-gap bound state. (b) and (c) show energies of mid-gap bound states
along axes at $\protect\lambda _{v}=0$ and $\protect\lambda _{R}=0$,
respectively. Regions marked by blue are for band states. It is seen that
energies of mid-gap bound states approach band edges when boundaries of the
TI phase are approached.}
\label{fig2}
\end{figure}

Using Eq. (\ref{1Dg}), the origin of the vacancy state is made clear. We
first note that the bulk Green's function in Eq. ({\ref{Ebound}) satisfies
\begin{equation}
g(\mathbf{0},\mathbf{0},E)=\sum_{k_{\zeta }}g(x_{\eta }=0,x_{\eta }^{\prime
}=0,k_{\zeta },E).  \label{integration}
\end{equation}%
Combining Eqs. (\ref{1Dg}) and (\ref{integration}), we find
\begin{equation}
g(\mathbf{0},\mathbf{0},E)=\left(
\begin{array}{cc}
w_{t}E-\mu _{t} & 0 \\
0 & w_{t}E-\mu _{t}%
\end{array}%
\right) ,  \label{g0}
\end{equation}%
where $w_{t}=\sum_{k_{\zeta }}w$, $\mu _{t}=\sum_{k_{\zeta }}w\mu _{\zeta }$
and we have made use the fact that the summation $\sum wk_{\zeta }$ vanishes
due to that $w(k_{\zeta },E)$ is an even function of $k_{\zeta }$. Eq. (\ref%
{g0}) then yields two midgap bound states with a degenerate energy at
\begin{equation}
E_{0}=\bar{\mu}_{\zeta }\equiv \frac{\sum_{k_{\zeta }}w\mu _{\zeta }}{%
\sum_{k_{\zeta }}w}.
\end{equation}%
Here $\bar{\mu}_{\zeta }$ is the weighted average of the intersecting energy for two helical
modes and hence it must lie inside the gap
\begin{equation}
|\mu _{\zeta }|<\Delta _{\zeta },  \label{vacancystate}
\end{equation}%
where $\Delta _{\zeta }$ is the gap along $\zeta $-direction. In the TI
phase, since there are edge states inside the energy gap for all
orientations of edges, $E$ must satisfy Eq. (\ref{vacancystate}) for all
possible $\zeta $, including the minimum of $\Delta _{\zeta }$, which is the
energy of the system. Hence one concludes that in consistent with the
Kramers degeneracy theorem, there must be a pair of degenerate vacancy
states inside the energy gap in the TI phase. }

To verify the above conclusion, we evaluate energies of midgap bound states
induced by a single vacancy through solving Eq. (\ref{Ebound}) numerically.
Figure~{\ref{fig2} shows energies of midgap bound states induced by a single
vacancy. It is seen that generally the midgap bound energy $E_{0}$ is not
fixed to zero and depends on values of parameters $\lambda _{R}$ and $%
\lambda _{SO}$. As one approaches boundaries of the TI phase, $E_{0}$ moves
towards edges of band gap. In the trivial insulator phase, the midgap bound
state merges into the bulk band and become a resonant state. Hence we
conclude that only the TI phase supports midgap bound states. }

\section{Impurity band and its effects on topological characterization}

\begin{figure}[tbp]
\begin{center}
\includegraphics[height=3.2in,width=3.2in] {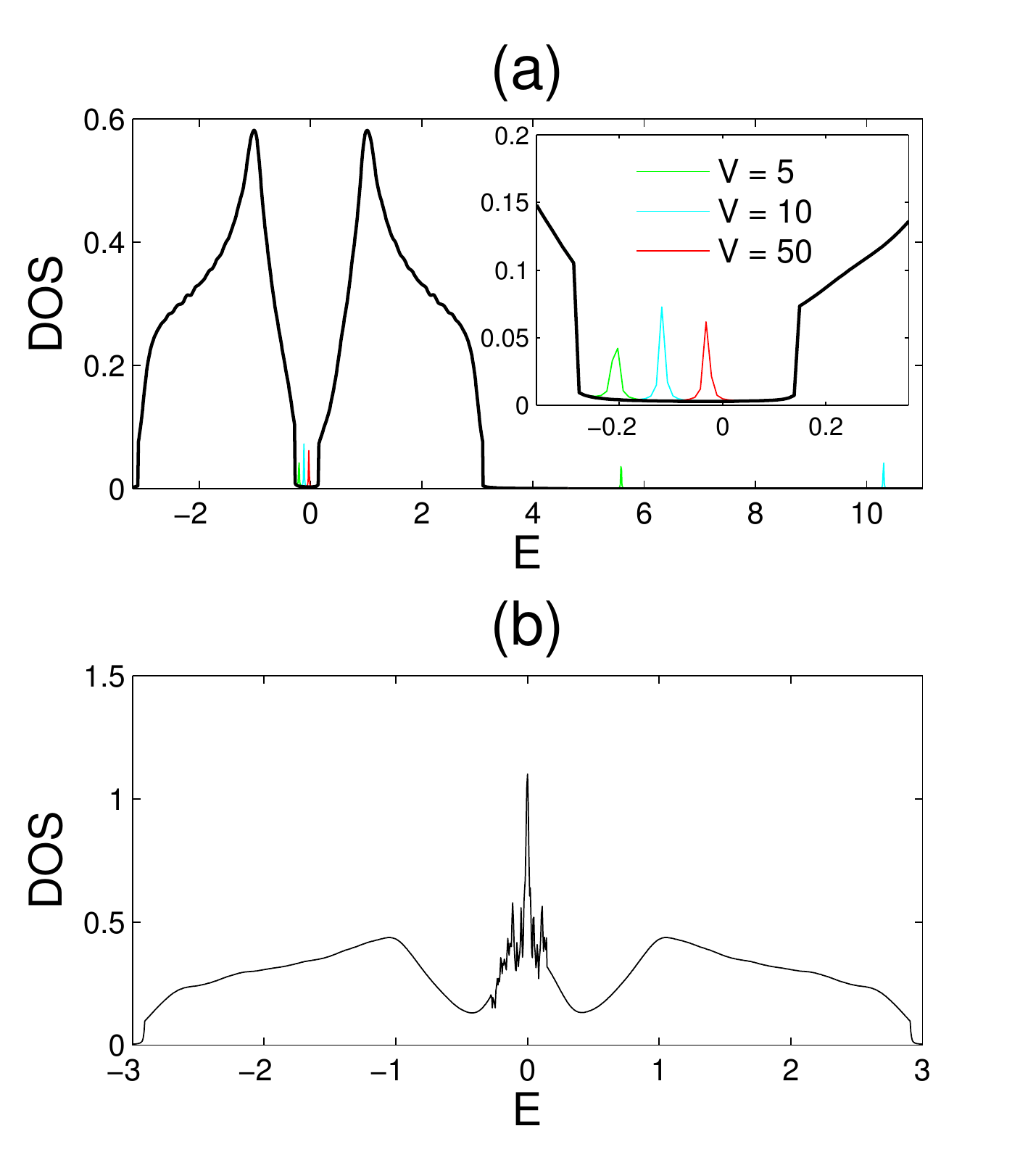}
\end{center}
\caption{(Color online) Typical density of states for a single impurity (a)
and for many vacancies (b). Here energies are in unit of $t$, $\protect%
\lambda _{R}=0.075$, and $\protect\lambda _{SO}=0.18\protect\sqrt{3}$. The
vacancy density in (b) is $1/9$. It is seen that an impurity band emerges
for finite density of vacancies, in agreement with the observation of Ref.
\protect\cite{Ando}.}
\label{fig3}
\end{figure}

In this section, we investigate effects of vacancies on Z$_{2}$ topological
order. For this purpose, we shall first investigate the distribution of
states induced by vacancies. In Fig.~\ref{fig3}(a), we show typical changes
of density of states for the system in the presence of a single impurity
when $V$ increases from $0$ to a very large magnitude. It is seen that two
midgap bound states degenerate in energy emerge from the band edge and
settles at a fixed energy inside the energy gap when $V$ approaches $\infty $%
. Furthermore, it is shown that whenever a midgap bound state is induced, a
high energy state at $E\sim V$ is also induced at the same time. The high
energy state is localized precisely at the impurity site and is pushed to
infinity when $V$ goes to $\infty $. Therefore, each vacancy induces two
midgap states degenerated at one midgap energy. When there are $M$ vacancies
at random positions, $M$ midgap energies are induced. Figure~\ref{fig3}(b)
shows that as the number of vacancies increases, these midgap bound states
start to form an impurity band inside the gap. The appearance of an impurity
band is in agreement with the observation made in Ref. \cite{Ando}. These
midgap states are localized states. Note that since the system is composed by spin-1/2
electrons with time-reversal symmetry and without spin-rotation symmetry,
disordered topological insulators belong to symplectic symmetry class \cite%
{Furusaki}. The level statistics of midgap energies is thus characterized by
the so-called Gaussian sympletic ensemble \cite{Mehta}. In particular, for a
Gaussian sympletic ensemble, the level-spacing distribution follows
Wigner-Dyson type distribution \cite{Mehta}. Hence for midgap energies in the impurity
band, the distribution of level spacings also follows the Wigner-Dyson
distribution.

From Fig.~\ref{fig3}(a), we have the following counting for level
distribution. In a honeycomb lattice with the number of unit cells being $N$%
, there are $2N$ lattice points and the total number of states is $4N$. If
there are $M$ vacancies, the distribution of energies is as follows
\begin{equation}
\mathrm{{Level~~counting}\left\{
\begin{array}{l}
{valence~~band:~~}2N-M \\
{impurity~~band:~~}M \\
{conduction~~band:~~}2N-M \\
{high~~energy~~states:~~}M.%
\end{array}%
\right. }
\end{equation}

By using the above level distribution, one can characterize the Z$_{2}$
topological index for each band. For this purpose, we first note that the
defining property of the non-trivial Z$_{2}$ topological index is the
Quantum spin Hall effect (QSHE). Therefore, the Z$_{2}$ topological index
can be evaluated by computing quantum spin Hall conductance (QSHC).
Computing the Z$_{2}$ index by evaluating QSHC is particularly useful when
there is no translational invariance in the presence of vacancies and one
can not define electronic Bloch states. To calculate QSHC, generally one
imposes spin-dependent twisted boundary conditions \cite{Sheng107} as
\begin{eqnarray}
&&c_{j+N_{x}\hat{x}}^{\alpha }=e^{i\theta _{x}^{\alpha }}c_{j}^{\alpha },
\nonumber \\
&&c_{j+N_{y}\hat{y}}^{\beta }=e^{i\theta _{y}^{\beta }}c_{j}^{\beta }.
\end{eqnarray}%
Here $\alpha ,\beta =\pm 1$ are the spin indices that represent the spin
state of $\uparrow $ and $\downarrow $ respectively. $j$ specifies the site,
$N_{x}$ and $N_{y}$ are numbers of sites along $x$ and $y$ directions
respectively [refer to Fig.~\ref{fig1}(a)], and $\theta _{x}^{\alpha }$ ($%
\theta _{y}^{\beta }$) are phases acquired whenever an electron with spin
component $\alpha $ ($\beta $) goes across the boundary along $x$ ($y$)
directions respectively. Both $\theta _{x}^{\alpha }$ and $\theta
_{y}^{\beta }$ are in the range $[0,2\pi ]$. Following Ref.~\cite{Sheng107},
when the degeneracy between spin state $+1$ and $-1$ is lifted, the Hall
conductances are generally represented as $\sigma _{xy}^{\alpha \beta }$
which characterizes the Hall voltage in the
$x$ direction due to electrons of spins component $\alpha$ 
when the spin-polarized current with spin component $\beta$ flows in the $y$ direction. 
The topological Chern number that is associated
with $\sigma _{xy}^{\alpha \beta }$ is thus defined as \cite{Sheng107}
\begin{equation}
C_{\alpha \beta }=\frac{i}{4\pi }\int \int d\theta _{x}^{\alpha }d\theta
_{y}^{\beta }\left[ \langle \frac{\partial \psi }{\partial \theta
_{x}^{\alpha }}|\frac{\partial \psi }{\partial \theta _{y}^{\beta }}\rangle
-\langle \frac{\partial \psi }{\partial \theta _{y}^{\beta }}|\frac{\partial
\psi }{\partial \theta _{x}^{\alpha }}\rangle \right] ,  \label{cab}
\end{equation}%
where $\psi $ is the many-particle ground state wavefunction of the system. $%
C_{\alpha \beta }$ can be collectively represented as a $2\times 2$ matrix,
representing the non-Abelian nature of the Hall conductance. For the quantum
Hall effect (QHE), it is entirely due to charges. The Hall conductance of
QHE is given by $\sigma _{H}=\sum_{\alpha ,\beta }\sigma _{xy}^{\alpha \beta
}$. Hence it is clear that $\sum_{\alpha ,\beta }C_{\alpha \beta }$ is the
corresponding Chern number associated with QHE. On the other hand, the QSHE
is defined as the difference of Hall conductances between spin $\uparrow $
and $\downarrow $. The QSHC $\sigma _{sH}$ is then given by
\begin{equation}
\sigma _{sH}=\sum_{\beta }\sigma _{xy}^{+\beta }-\sum_{\beta }\sigma
_{xy}^{-\beta }=\sum_{\alpha \beta }\alpha \sigma _{xy}^{\alpha \beta }.
\end{equation}%
As a result, the corresponding spin Chern number that is associated with
QSHE is given by
\begin{equation}
C_{s}=\sum_{\alpha \beta }\alpha C_{\alpha \beta }.  \label{sChern1}
\end{equation}%
From Eqs. (\ref{cab}) and (\ref{sChern1}), it is seen that the minus sign
associated with down spin can be absorbed into the twisted phase $\theta
_{x}^{\downarrow }$ so that one can set $\theta _{x}^{\downarrow }=-\theta
_{x}^{\uparrow }$. Therefore, to focus on the spin Chern number, one sets $%
\theta _{y}^{\uparrow }=\theta _{y}^{\downarrow }=\theta _{y}$ and $\theta
_{x}^{\uparrow }=-\theta _{x}^{\downarrow }=\theta _{x}$ and imposes the
spin-dependent twisted boundary conditions \cite{Sheng107,Hatsugai2} as
follows
\begin{eqnarray}
&&c_{j+N_{x}\hat{x}}=e^{i\theta _{x}\sigma _{z}}c_{j},  \nonumber \\
&&c_{j+N_{y}\hat{y}}=e^{i\theta _{y}}c_{j},
\end{eqnarray}%
where $\sigma _{z}$ is the $z$ component of the Pauli matrices and both $%
\theta _{x}$ and $\theta _{y}$ are in the range $[0,2\pi ]$. The gauge
potential $\Phi $ imposed by twisted boundary conditions is
\begin{equation}
\Phi (i_{x},i_{y})=\sigma _{z}\frac{\theta _{x}i_{x}}{N_{x}}+\frac{\theta
_{y}i_{y}}{N_{y}}.
\end{equation}%
Due to the presence of $\sigma _{z}$, $\Phi $ does not commute with $H$,
which reflects the non-Abelian nature of the problem. Note that the spin
Chern number computed by the spin-dependent twisted boundary conditions has
been rigorously shown to be equivalent to the Z$_{2}$ topological index \cite%
{Hatsugai2,Vanderbilt}.

Following Refs.~\cite{Hatsugai1} and \cite{Hatsugai2}, using the energy
eigenkets of $H$ with spin-dependent twisted boundary conditions, one can
compute the spin Chern number which yields the same classification as that
of Z$_{2}$ topological order. For this purpose, we discretize the space $%
[0,2\pi ]\times \lbrack 0,2\pi ]$ into $M\times M$ mesh points so that a
general twisted boundary condition is represented by $\theta \equiv (\theta
_{p},\theta _{q})$ with $\theta _{p}=2\pi p/M$, $\theta _{q}=2\pi q/M$ and $%
p,q=0,1,2,3,\cdots ,M-1$. For each $\theta $, since hopping across
boundaries depends on $\theta $, the Hamiltonian depends on $\theta $, $%
H(\theta )$, with eigenstates being $|n(\theta )\rangle $. For each bond
that connects $\theta $ and $\theta +\delta \theta _{i}\hat{\imath}$ with $%
\delta \theta _{i}=2\pi /M$ and $i=x$ or $y$, a non-Abelian link variable $%
U_{i}(\theta )$ is defined as the overlap matrix
\begin{equation}
U_{i}(\theta)_{mn}=\langle m(\theta )|n(\theta +\delta \theta _{i}\hat{\imath%
} )\rangle,  \label{link}
\end{equation}
where $n$ and $m$ are indices of energy bands that are included for
computing the spin Chern number. From the link variable $U_{i}(\theta )$, a
U(1) link $u_{i}(\theta )$ can be formed
\begin{equation}
u_{i}(\theta )=\det U_{i}(\theta )/|\det U_{i}(\theta )|.
\end{equation}
One can then find the lattice field strength $F_{xy}$ for each plaquette in
the $(\theta _{x},\theta _{y})$ lattice by computing
\begin{eqnarray}
F_{xy}(\theta ) &=&\ln \left[ u_{x}(\theta )u_{y}(\theta +\delta \theta _{x}%
\hat{x})\right. \\
&&\left. u_{x}^{-1}(\theta +\delta \theta _{x}\hat{x}+\delta \theta _{y}\hat{%
y})u_{y}^{-1}(\theta +\delta \theta _{y}\hat{y})\right] ,  \nonumber
\end{eqnarray}%
where the principle branch of logarithm is taken. The spin Chern number is
the summation of $F_{xy}$ over all plaquettes given by
\begin{equation}
C_{s}=\frac{1}{2\pi i}\sum_{\theta }F_{xy}(\theta ).  \label{sChern}
\end{equation}%
The spin Chern number thus obtained is always an integer. Furthermore, it
has the advantage of being accurate even when the computation is done with
small sizes of honeycomb lattices \cite{Hatsugai1,Hatsugai2}.

To characterize TI with vacancies at random positions, we compute $C_s$ for
valence band, impurity band and conduction band. For a given density of
vacancies, it is found that depending on the configuration of positions for
vacancies, the spin Chern number of each band can be either $2$ or $0$. As a
result, one has to perform average of $C_s$ over configurations of
vacancies. However, we find that as long as there is no overlap between
these bands, $C_s$ of the impurity band always vanishes. Hence $C_s$ of the
conduction band is always opposite to that of the valence band.

To confirm the validity of computed $C_{s}$ based on Eq. ({\ref{sChern}),
for a given vacancy configuration, we examine the bulk-edge correspondence
by computing the energy spectrum for honeycomb ribbons with the armchair
edges. To this end, it is convenient to compute the spectral function $A(
\vec{k},\omega )$ defined by
\begin{equation}
A(\vec{k},\omega )=-\frac{1}{N\pi }\rm{Im}\left\{
\sum_{j=1}^{4}\sum_{n,m=1}^{4N}\frac{|\langle n|\psi (\vec{k},j,m)\rangle
|^{2}}{(\omega -E_{n})+i\Gamma }\right\} ,  \label{spectral function}
\end{equation}%
where $\Gamma $ is an energy scale characterizing energy resolution, $%
|n\rangle $ is the \textit{n}-th energy eigenstate of the impure system with
eigen-energy $E_{n}$, and $|\psi (\vec{k},j,m)\rangle $ is the energy
eigenstate in the absence of vacancy with $\vec{k}$ being the wave vector
after Fourier transformation. Figure~\ref{fig4} shows the spectral function $A$ of a honeycomb ribbon with armchair
edges in the presence of vacancies for different vacancy densities. Here the
listed spin Chern numbers are computed by using Eq.(\ref{sChern}) based on
twisted boundary conditions. The armchair edges are along the $x$ direction
so we have the conserved wave vector $k_{x}$. It is seen that for small
vacancy densities when $C_{s}$ does not vanish for both valence and
conduction bands, edges states exist and coexist with the impurity band at
the center. Furthermore, in consistent with the bulk-edge correspondence,
edge states reside between the impurity band and the valence band or the
conduction band. As the density of vacancies increases, the strength of the
impurity band increases and eventually }$C_{s}${\ for all energy bands
vanish. In this case, as shown in Fig.~\ref{fig4}(c), edge states diminish
as well.
\begin{figure*}[tbp]
\includegraphics[width=18cm ]{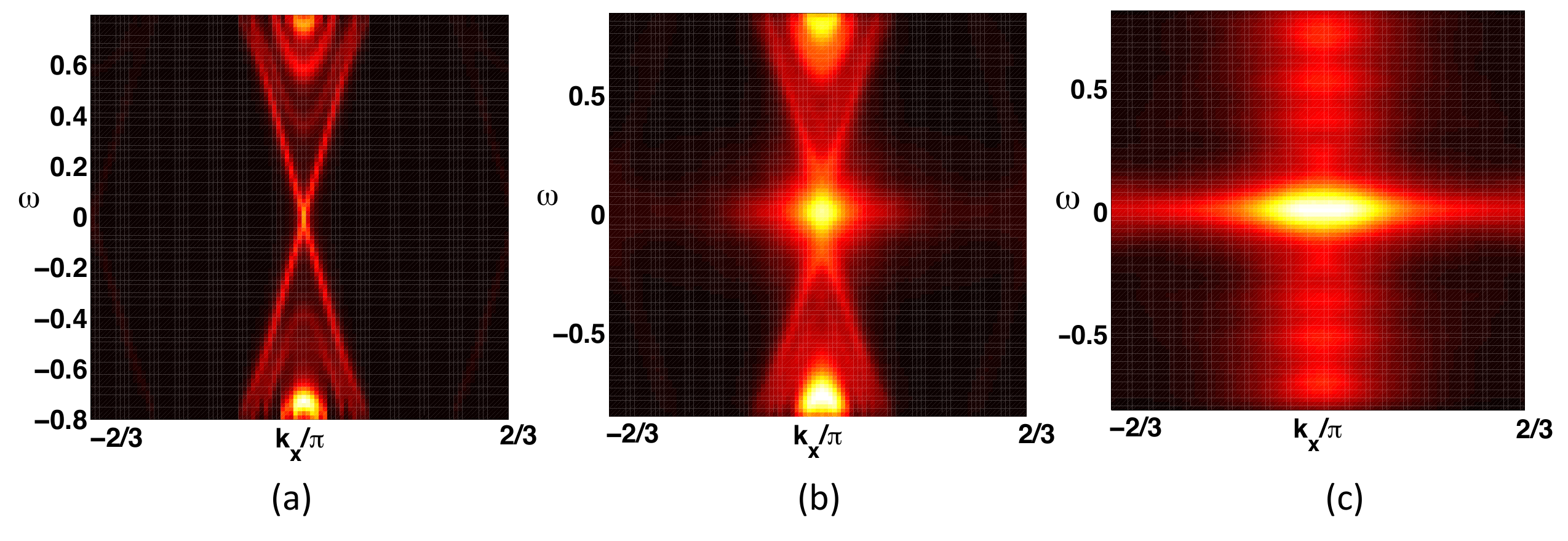}
\caption{(Color online) Coexistence of edge states and the impurity band
exhibited in the spectral function $A(k_{x},\protect\omega )$ of a honeycomb
ribbon with armchair edges for different vacancy densities: (a) 0\% ($C_{s}=2
$ for the valence band, $C_{s}=-2$ for the conduction band) (b) 3\% ($C_{s}=2
$ for valence band, $C_{s}=0$ for the impurity band, $C_{s}=-2$ for the
conduction band) (c) 30\% ($C_{s}=0$ for all energy bands). Here $k_{x}$ is
the wave vector along armchair edges, $\Gamma =0.01$, $\protect\lambda %
_{R}=0.075$, $\protect\lambda _{SO}=0.18\protect\sqrt{3}$, and the size of
the honeycomb ribbon is $100\times 8$. Here the listed spin Chern numbers
are computed by using Eq.(\protect\ref{sChern}) based on twisted boundary
conditions.}
\label{fig4}
\end{figure*}
}

\begin{figure}[h]
\begin{center}
\includegraphics[height=2.3in,width=3.0in] {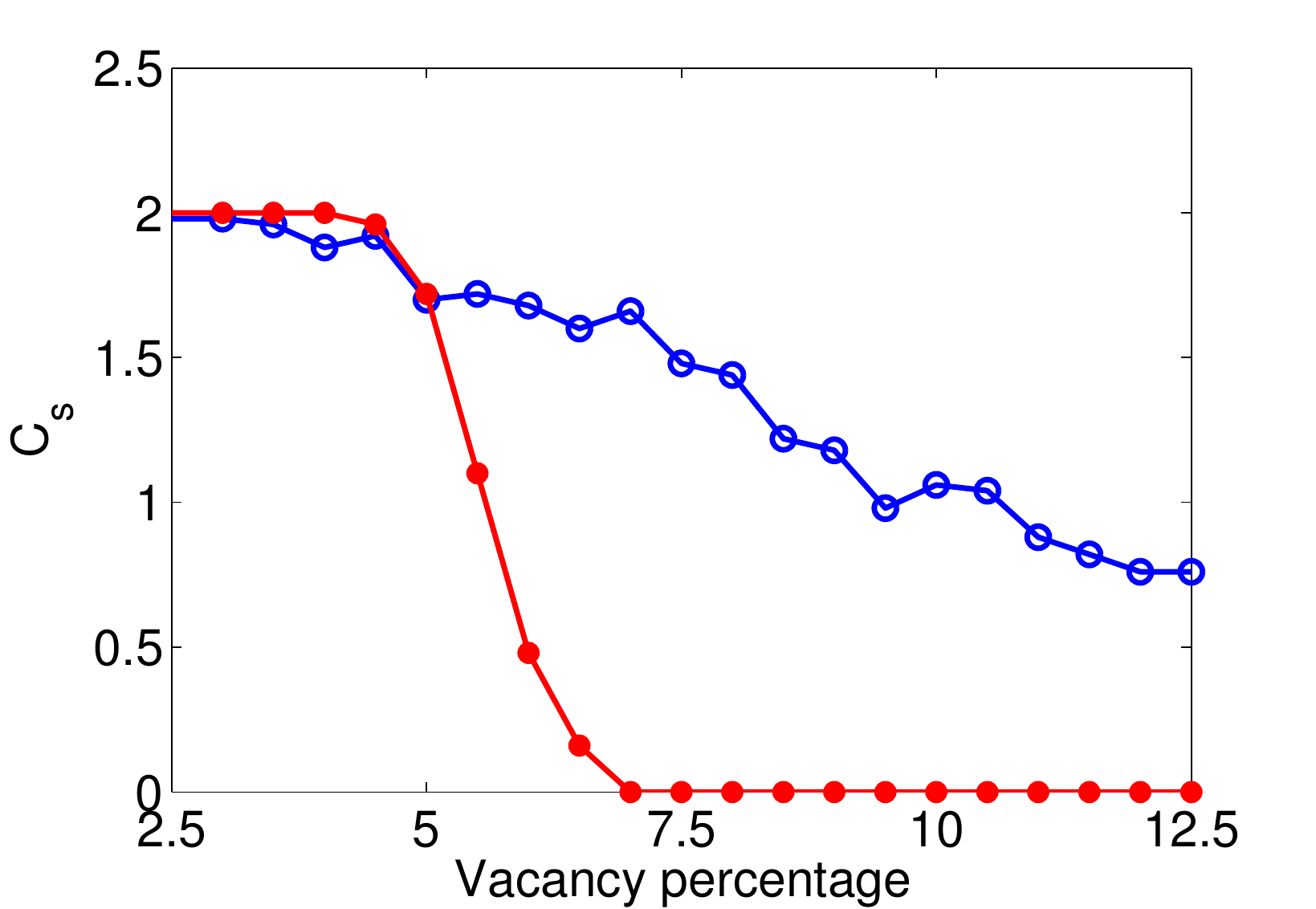}
\end{center}
\caption{(Color online) Phase transition of TI to a trivial insulator due to
the presence of vacancies at random positions. Here the averaged spin Chern
number ($C_{s}$) is computed for the valence band with vacancies being
located only at \textit{A} sites (solid circle) or equally distributed at
\textit{A} sites and \textit{B} sites (open circle). Parameters in the plot
are $M=10$, $\protect\lambda _{R}=0.075$, $\protect\lambda _{SO}=0.18\protect%
\sqrt{3}$, and the average is taken over $100$ vacancy configurations.}
\label{fig5}
\end{figure}

As indicated in the above, to explore the nature of transition, it is
necessary to compute the averaged spin Chern number. In Fig.~{\ref{fig5}, we
show the averaged spin Chern number for the valence band of TIs in the
presence vacancies which positions are only at \textit{A} sites or
distribute equally at \textit{A} sites and \textit{B} sites. The validity of
computations based on Eq.(\ref{sChern}) is also checked by direct
computation of the corresponding systems with open boundaries. It is seen
that a phase transition from TI to a trivial insulator occurs as density of
vacancies increases. However, depending on positions of vacancies, the
transition can be either first order when vacancies are randomly chosen at
\textit{A} sites or continuous when vacancies are randomly chosen equally at
\textit{A} sites or \textit{B} sites. As we shall explore in the following,
the mechanism that causes different transition behaviors lies in the speed
of the Dirac cones being depleted. Since the formation of Dirac cones
involves with both \textit{A} sites and \textit{B} sites, removing points in
\textit{A} sites and \textit{B} sites simultaneously will start to deplete
the Dirac cone and the depletion is proportional to the number of points
being removed. As a result, the transition is continuous when vacancies are
distributed equally at \textit{A} sites and \textit{B} sites. On the other
hand, removing points solely at \textit{A} sites does not deplete the Dirac
cone immediately at first until when too many A points are removed, the
bi-partite nature is destroyed. At that point, further increase of vacancies
induces a discontinuous transition to a trivial insulator with $C_{s}=0$. }

\begin{figure*}[tbp]
\includegraphics[width=15cm ]{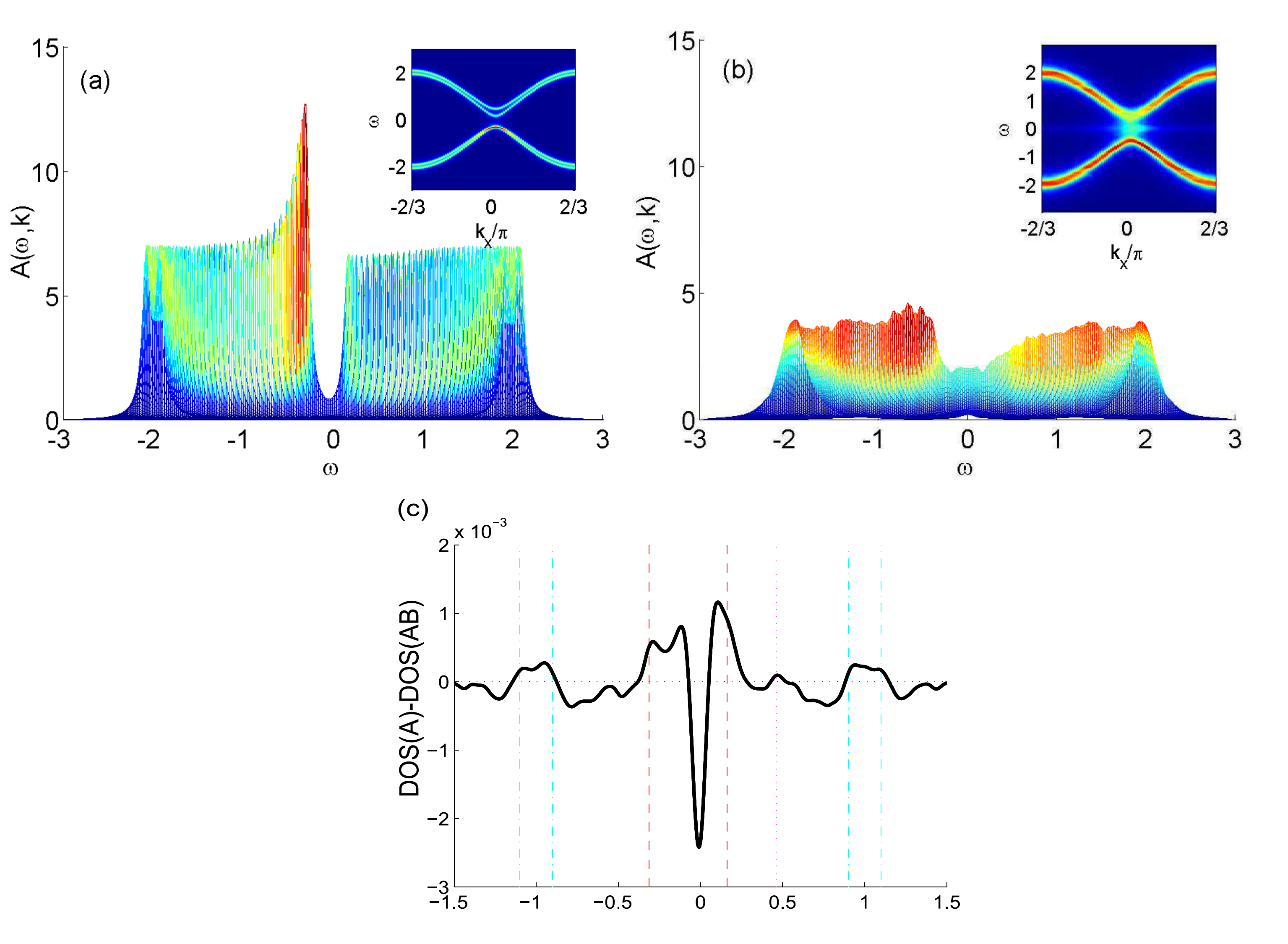}
\caption{(Color online) Depletion of Dirac cones exhibited in $A(k,\protect%
\omega )$ and the density of states. Here (a) and (b) show the side view of $%
A(k,\protect\omega )$ along $\protect\omega $ axis for a honeycomb lattice
of size 100 $\times $ 8 with different number of vacancies: (a) 0\% (b)
6.26\%. A sharp peak at Dirac point is clearly revealed in (a) but is
depleted quickly when vacancies are introduced, as shown in (b). (c)
Difference of densities of states for 20 vacancies in a 50$\times $50
honeycomb lattice with vacancies being solely distributed at \textit{A}
sites, DOS(A), and vacancies being distributed equally at \textit{A} sites
and \textit{B} sites, DOS(AB). Here $\protect\lambda _{R}=0.075$, and $%
\protect\lambda _{SO}=0.18\protect\sqrt{3}$. Dashed and dotted lines
indicate energy of gaps at Dirac points, while dot-dashed lines indicate
energies at M points. It is seen that the Dirac cones are less depleted when
vacancies are solely distributed at \textit{A} sites.}
\label{fig6}
\end{figure*}

To explore the mechanism of how Z$_{2}$ topological order is destroyed by
vacancies, we first note that near the phase boundary when the energy gap
vanishes, the dominant contribution of the spin Chern number is from $k$
points near the Dirac points due to the singularity induced by band touching
of conduction and valence bands. It is known that the Dirac cones continue
to dominate the contribution of Z$_{2}$ topological order even when the
system is away from the phase boundary \cite{Ezawa}. Therefore, it is
essential to examine how Dirac cones evolve with the introduction of
vacancies. For this purpose, we examine the spectral function $A(\vec{k}%
,\omega )$ near Dirac cones using the periodic boundary conditions so that
the spectral weight only reflects effects due to vacancies not due to edge
states. In Fig.~\ref{fig6}(a), we show side view of $A(\vec{k},\omega )$ of
a honeycomb lattice in the TI phase with a size of 100$\times $8. A clear
peak located right at the Dirac point is exhibited. The peak that reflects
the spectral weight near Dirac cones is depleted quickly once vacancies are
introduced, as shown in Fig.~\ref{fig6}(b). Here the number of vacancies is
90, \textit{i.e.}, density of vacancies is 6.25\%. Clearly, it shows that
the diminishing of $C_{s}$ correlates with the depletion of Dirac cones. To
further differentiate effects due to distribution of vacancies on different
sub-lattices, we compute the density of states for 20 vacancies in a 50$%
\times $50 honeycomb lattice with vacancies being solely distributed at
\textit{A} sites, denoted by DOS(A), and vacancies be distributed equally at
\textit{A} sites and \textit{B} sites, denoted as DOS(AB). The difference of
densities of states is shown in Fig.~\ref{fig6}(c). Clearly, the difference,
DOS(A)-DOS(AB), is pronounced at several energies. Here dashed and dotted
lines indicate energies of gaps at Dirac points (K point), while dot-dashed
lines indicate energies at M points. It is seen that removing points solely
at \textit{A} sites tends to keep a higher spectral weight at symmetry
points (K and M points) in $k$ space. As a result, the Dirac cones are less
depleted when vacancies are distributed solely at \textit{A} sites. The
analyses shown here show that direct depletion of the Dirac cones is the
main mechanism for transition from the TIs to trivial insulators for low
concentrations of vacancies. For higher concentration of vacancies, as shown
in Fig.~\ref{fig4}, the impurity band starts to overlap with energy bands
and destructs of Z2 topological order eventually.

\section{Discussion and conclusion}

In summary, we have shown that vacancies in the TI phase of the Kane-Mele
model always induce midgap bound states. These midgap bound states result
from curving edge states into the surroundings of vacancies. The same
reasoning can be generalized to investigate point vacancies or line
vacancies in 3D TIs. In this case, the T-matrix still obeys Eq. (\ref%
{Tmatrix}). By introducing a plane of vacancies passing through $\vec{r}=%
\mathbf{0}$ that cuts infinite 3D lattice into two semi-infinite 3D lattices
and following similar considerations that lead to Eq. (\ref{integration}),
one arrives at
\begin{equation}
g(\mathbf{0},\mathbf{0},E)=\sum_{k_{\alpha },k_{\beta }}g(x_{\gamma
}=0,x_{\gamma }^{\prime }=0,k_{\alpha },k_{\beta },E),  \label{integration3D}
\end{equation}%
where $k_{\alpha }$ and $k_{\beta }$ are wave vectors in two orthogonal
directions of the plane with vacancies and $x_{\gamma }$ is the coordinate
of the axis in perpendicular to the plane. For 3D TIs, since there are
midgap surface states characterized by spectrum $\pm v_{\parallel
}k_{\parallel }+\mu _{\parallel }$ with $k_{\parallel }=\sqrt{k_{\alpha
}^{2}+k_{\beta }^{2}}$, the same reasoning leads to the conclusion that in
the low energy sector, $g$ must take the following form
\begin{eqnarray}
&&g(0_{\gamma },0_{\gamma },k_{\alpha },k_{\beta },E)=  \nonumber \\
&&w(k_{\alpha },k_{\beta },E)\left[
\begin{array}{cc}
E-\mu _{\parallel } & v_{\parallel }(k_{\alpha }-ik_{\beta }) \\
v_{\parallel }(k_{\alpha }+ik_{\beta }) & E-\mu _{\parallel }%
\end{array}%
\right] ,  \label{2Dg}
\end{eqnarray}%
where $v_{\parallel }$ is the speed of electrons in surface states, $\mu
_{\parallel }$ is the energy of the Dirac point, and $w(k_{\alpha },k_{\beta
},E)$ is the weight associated with $k_{\parallel }$ and is a smooth
function of $E$. Hence we arrive at a conclusion similar to Eq. (\ref%
{vacancystate}) with $\Delta _{\zeta }$ being replaced by the gap $\Delta
_{\parallel }$ of $\vec{k}$ that lies in directions perpendicular to $\gamma
$. As a result, one also concludes that there must be a pair of degenerated
states inside the energy gap for a single vacancy in 3D topological
insulators. This conclusion is in agreement with direct numerical
simulations investigated in Ref. \cite{Balatsky}. Similarly, for a singe
line of vacancies along $\delta $ direction, performing a partial Fourier
transform on the Hamiltonian along $\delta $ reduces the problem to a 2D
problem with a point vacancy. Hence there must be a pair of degenerated
states inside the energy gap for every $k_{\delta }$ mode along each line of
vacancies in 3D TIs.

In addition to showing that midgap bound states must arise in the presence
of a single vacancy, we also show that an impurity band emerges when the
number of vacancies increases. The impurity band mixes directly with edge
states. However, the impurity band always has trivial topological structure
and the Z $_{2}$ topological order persists in both the conduction band and
the valence band. Hence edge states exist between the impurity band and
perturbed conduction or valence band. The mechanism behind the transition
from TI to a trivial insulator due to vacancies is shown to be resulted from
the depletion of Dirac cones. Furthermore, due to different speeds of
depleting Dirac cones, the transition can be either first order when
vacancies are randomly chosen at \textit{A} sites or continuous when
vacancies are randomly chosen equally at \textit{A} sites or \textit{B}
sites.

While so far in this work we only consider effects of vacancies, our results
are not restricted to the limit $V\rightarrow \infty $. For finite and large
$V$, Eq. (\ref{Ebound}) is modified with a correction of $O(1/V)$. Hence
midgap bound states induced by vacancies are shifted by the order of $O(1/V)$%
, in agreement with results shown in Fig.~\ref{fig3}. The phase boundary for
midgap bound states shown in Fig.~\ref{fig2} then remains the same. As a
result, the impurity band is also shifted by a similar amount. The majority
of our results remains valid and provides useful characterization of
electronic states induced by point defects in TIs.

\begin{acknowledgments}
We thank Profs. Ming-Che Chang and Sungkit Yip for useful discussions. This
work was supported by the National Science Council of Taiwan.
\end{acknowledgments}

\end{document}